\documentclass[conference]{IEEEtran}
\IEEEoverridecommandlockouts
\usepackage{cite}
\usepackage{amsmath,amssymb,amsfonts}
\usepackage{algorithmic}
\usepackage{graphicx}
\usepackage{textcomp}
\usepackage{xcolor}
\usepackage{array}
\usepackage{booktabs}
\usepackage{gensymb}
\usepackage{caption} 
\usepackage{hyperref}

\def\BibTeX{{\rm B\kern-.05em{\sc i\kern-.025em b}\kern-.08em
    T\kern-.1667em\lower.7ex\hbox{E}\kern-.125emX}}
\begin{document}

\title{\textsc{MuViS}: Multimodal Virtual Sensing Benchmark
\thanks{Equal contribution: \{j.u.brandt, n.c.puetz\}@liacs.leidenuniv.nl}
}

\author{
    \IEEEauthorblockN{
        Jens U. Brandt\textsuperscript{*,a,b}, 
        Noah C. Puetz\textsuperscript{*,a,b}, 
        Jobel Jose George\textsuperscript{a}, 
        Niharika Vinay Kumar\textsuperscript{a}, \\
        Elena Raponi\textsuperscript{b},
        Marc Hilbert\textsuperscript{c,b},
        Thomas Bäck\textsuperscript{b}, and 
        Thomas Bartz-Beielstein\textsuperscript{a}
    }
    \IEEEauthorblockA{
        \textsuperscript{a}TH Köln, Germany. 
        \textsuperscript{b}Leiden University, Netherlands. 
        \textsuperscript{c}Toyota Racing, Germany.
    }
}

\maketitle

\begin{abstract}
Virtual sensing aims to infer hard-to-measure quantities from accessible measurements and is central to perception and control in physical systems. Despite rapid progress from first-principle and hybrid models to modern data-driven methods research remains siloed, leaving no established default approach that transfers across processes, modalities, and sensing configurations. We introduce \textsc{MuViS}, a domain-agnostic benchmarking suite for multimodal virtual sensing that consolidates diverse datasets into a unified interface for standardized preprocessing and evaluation. Using this framework, we benchmark established approaches spanning gradient-boosted decision trees and deep neural network (NN) architectures, and show that none of these provides a universal advantage, underscoring the need for generalizable virtual sensing architectures. \textsc{MuViS} is released as an open-source, extensible platform for reproducible comparison and future integration of new datasets and model classes.
\end{abstract} 


\section{Introduction}
\label{sec:Intro}
Accurate monitoring of internal system states is fundamental for autonomous systems, industrial automation, and structural health monitoring. Sensors play a crucial role by translating the physical world into digital signals that can be processed, stored, and acted upon. Their importance continues to increase as modern engineering paradigms increasingly rely on dense, high-frequency instrumentation to enable closed-loop decision making and continuous system supervision \cite{fraden_handbook_2016}.
In particular, trends such as digital twins \cite{glaessgen_digital_2012}
, which require persistent, data-driven synchronization between a physical asset and its digital counterpart, and emerging notions of physical AI \cite{brooks_intelligence_1991} further amplify the need for reliable sensing and state inference.

Virtual sensing (also referred to as soft sensing) addresses the mismatch between the variables one \emph{wants} to measure and those one \emph{can} measure precisely. In many applications, the target of interest may be expensive, delayed, intrusive, or difficult to instrument directly, while related secondary variables are readily available through standard sensors. By fusing such sensor inputs, virtual sensors can estimate latent or otherwise hard-to-measure process variables and enable continuous monitoring without additional hardware \cite{martin_virtual_2021}. 

\begin{figure}[h]
\centerline{\includegraphics[width=0.45\textwidth]{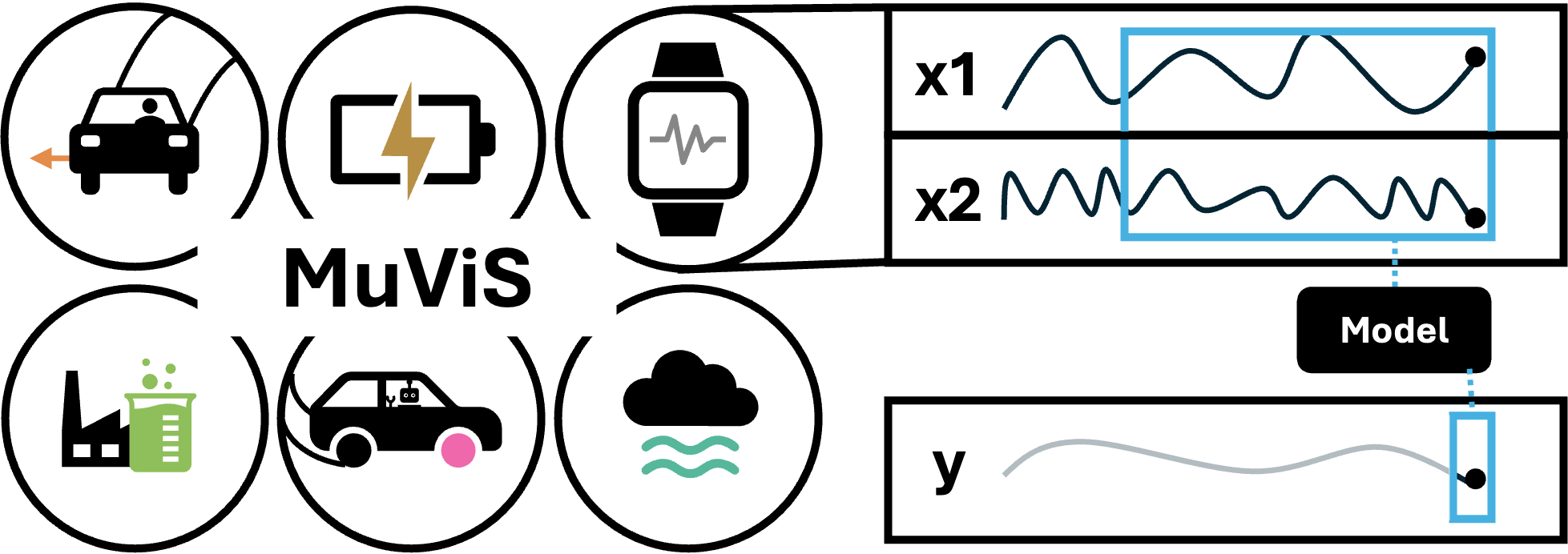}}
\caption{We evaluate standard ML architectures across diverse sensing domains, where models must map multimodal time-series inputs ($x1, x2$) to scalar virtual measurements ($y$).}
\label{fig:muvis_overview}
\vspace{-17pt}
\end{figure}

Following Martin et al. \cite{martin_virtual_2021}, we consider a physical sensor to be a technical device that reacts to a physical stimulus and outputs a regularly sampled signal. A virtual sensor, in contrast, is a software-defined sensing entity that produces a sensor-like measurement by combining, aggregating, and transforming signals obtained from physical transducers and, potentially, other virtual sources \cite{martin_virtual_2021}. The basic concept of virtual sensors dates back at least to Muir \cite{muir_virtual_1990}, who emphasizes that virtual sensing yields a measurement that appears to the user as if a physical sensor measured the desired parameter directly.

Virtual sensing methods include mechanistic \emph{white-box} models, data-driven \emph{black-box} models, and \emph{gray-box} hybrids that combine both \cite{chen_dynamic_2020}. While white/gray-box approaches can be preferable when reliable physics and expert knowledge are available \cite{chen_cognitive_2014}, many virtual sensing tasks involve multimodal, nonlinear, high-dimensional signals where faithful mechanistic models are unavailable or impractical \cite{reiss_deep_2019}. Moreover, for many deployments,
it is desirable to enable rapid prototyping and reuse without requiring substantial domain expertise,
which motivates our focus on \emph{black-box} models \cite{jiang_review_2021}. Despite progress in data-driven virtual sensing, evaluations remain application-centric, making it unclear how architectures transfer across disparate physical phenomena and modality compositions \cite{kadlec_soft_2009,jiang_review_2021}. Unlike computer vision, where broadly useful inductive biases (e.g., convolution and pooling) have led to comparatively standardized model families \cite{cohen_inductive_2017}, virtual sensing lacks an established ``default'' architecture that is consistently effective across heterogeneous modalities and sampling regimes. 

We address this gap by introducing \textsc{MuViS} (Multimodal Virtual Sensing), a comprehensive, domain-agnostic multimodal virtual sensing benchmark suite for evaluating machine learning (ML) models across diverse physical systems (Fig.~\ref{fig:muvis_overview}). By aggregating datasets from various sources, \textsc{MuViS} tests the predictive capabilities of modern architectures and goes beyond industry-specific evaluations.
Our contributions are three-fold: (i) We introduce unified datasets spanning \emph{six application domains} to enable standardized virtual sensing evaluation. (ii) We \emph{benchmark six representative models}, from gradient-boosted decision trees to deep NN architectures. (iii) We release an \emph{open-source, extensible framework} \cite{noauthor_noah-puetzmuvis_nodate} for standardized preprocessing and evaluation, enabling straightforward integration of new datasets and architectures.

\begin{figure*}[t]
\centerline{\includegraphics[width=1.0\textwidth]{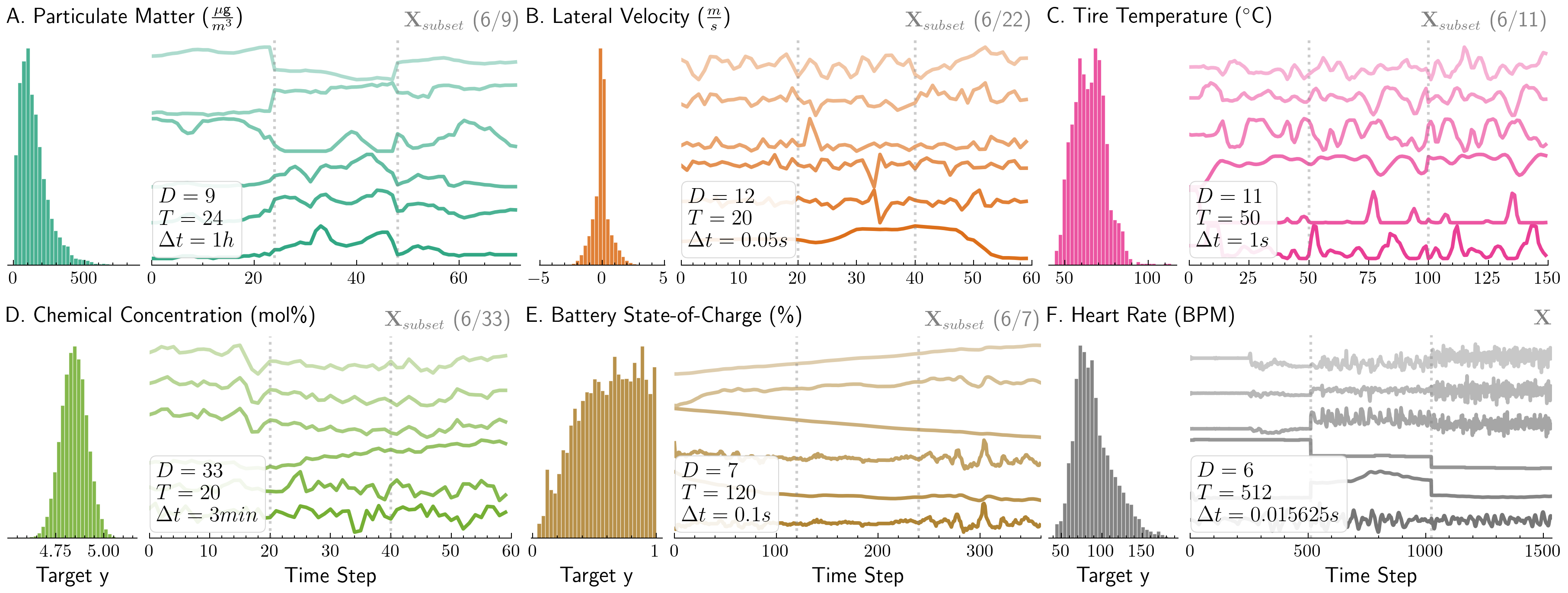}}
\caption{Overview of the six benchmark datasets. Each sub-panel displays a distinct virtual sensing task, showcasing the diversity in target distributions (left) and temporal feature characteristics (right). The collection spans varied feature dimensions ($D$), sequence lengths ($T$), and sampling intervals ($\Delta t$), reflecting the heterogeneous nature of real-world sensing applications.}
\label{fig_eyecatcher}
\vspace{-17pt}
\end{figure*}

\section{Defining Multimodal Virtual Sensing}
\label{sec:muvis}

Conceptually, virtual sensors operate on data that ultimately originates from physical sensing devices. Accordingly, we restrict the scope of \textsc{MuViS} to signals derived from instrumentation and exclude purely symbolic sources such as human annotations or sentiment labels. But, the boundary between ``physical'' and ``virtual'' sensing is not always sharp~\cite{martin_virtual_2021}: many physical sensors incorporate non-trivial analog/digital processing pipelines and calibration steps that translate a stimulus into an electrical signal and subsequently into an engineered measurement \cite{fraden_handbook_2016} therefore a strict interpretation could classify parts of the sensing chain as ``virtual``. In this work, we use the term \emph{virtual sensing} to denote the data-driven inferential step that maps available instrument signals to a hard-to-measure target variable, independent of signal conditioning performed inside sensor hardware.

We formulate \textsc{MuViS} as the intersection of virtual sensing and multimodal dynamic time-series learning \cite{mohapatra_maestro_2025}, with an interface to time-series extrinsic regression (TSER) \cite{tan_monash_2020}. Each sample  $\mathcal{X}_i \in X$ with $i\in\{1,\dots,N\}$ of is represented by a set of modality-specific sensor streams
\begin{equation}
    \mathcal{X}_i = (x_i^1,\ldots,x_i^M), 
    \qquad 
    x_i^j \in \mathbb{R}^{D^j \times T^j},
\end{equation}
where modality $j$ provides $D^j$ feature channels observed over $T^j$ time steps. We call the input multimodal if $M\geq 2$ and the modalities capture complementary views of the same underlying system. Following Mohapatra et al. \cite{mohapatra_maestro_2025}, we assume that modality streams are largely complementary (or ``semantically disjoint'') in the sense that no modality can be reliably reconstructed from another via a deterministic mapping:
\begin{equation}
\not\exists \phi_{jk}:\mathbb{R}^{D^{j}\times T^{j}}\rightarrow \mathbb{R}^{D^{k}\times T^{k}}
\end{equation}
such that
\begin{equation}
\phi_{jk}(x_i^{j}) \approx x_i^{k},
\ \forall j\neq k.
\end{equation}
Intuitively, each modality contributes information that is not fully redundant with the others, so predictive performance may depend on cross-modal interactions rather than on any single stream alone. This is conceptually aligned with \emph{cooperative} sensing notions, where multiple independent sensors collectively provide information that would not be available from an isolated view \cite{martin_virtual_2021}. In real systems, the set of available sensor modalities rarely constitutes a complete ``view'' of the sensed object or process: any instrumentation provides only partial information, and the specific subset of sensors deployed can differ across installations. To reflect this, we adopt the dynamic modality setting of Mohapatra et al. \cite{mohapatra_maestro_2025} which is closely aligned with Martin et al.`s \cite{martin_virtual_2021} notion of dynamic cooperative sensing. Each sample is associated with a subset of observed modality indices $S_i \subseteq \{1,\ldots,M\}$, and a model must operate on the observed collection $\{x_i^j \mid j\in S_i\}$. In this interpretation, $S_i$ captures variability in sensing configurations and context-dependent feature availability (e.g., different setups or operational conditions, changes in system composition or sampling rate), rather than being limited to accidental sensor failure.

In contrast to the classification target in multimodal dynamic time-series learning~\cite{mohapatra_maestro_2025},
\textsc{MuViS} is inherently a continuous estimation problem. We consider a target variable (or vector) $y_i(t_0)\in\mathbb{R}^d$ at a reference time $t_0$ and learn a virtual sensor $f:X_{S_i} \rightarrow \mathbb{R}^d$ that maps the available multimodal history to an estimate, i.e., $\hat{y}_i(t_0)=f\!\left(\{x_i^j \mid j\in S_i\}\right)$.

Operationally, $f$ may be instantiated as a function-to-function (or ``sequence-to-value'') mapping: given a temporal context window ending at $t_0$, the model predicts the target at $t_0$ using the recent dynamics of the observed modalities. Since modality streams can have different sequence lengths $T^j$, we set the common window length to $s \triangleq \max_{j\in S_i} T^j$ and treat shorter histories as left-padded, masked or upsampled.
\begin{equation}
    \hat{y}_i(t_0) = f\!\left(\{x_i^j[t]\}_{t=t_0-s+1}^{t_0,\; j\in S_i}\right)
\end{equation}
Importantly, this formulation does not correspond to forecasting future values; instead, the output is anchored at the reference time and interpreted as a sensor-like measurement that could, in principle, be produced by a physical sensor.

This formulation provides a unified lens for \textsc{MuViS}: across application domains, the learning problem can be cast as multimodal function-to-function regression, i.e., mapping a window of sensor measurements to a sensor-like target value at a reference time $t_0$. \textsc{MuViS} makes this shared structure explicit and enables evaluation across diverse sensing configurations. Domains differ in the number of modalities $M$, channel dimensionalities $D$, and sampling rates (and thus the chosen context length $T$), yet all conform to the same input--output characteristics. Consequently, the benchmarks introduced in the following sections focus on accurate continuous sensor-value estimation at $t_0$ and the model's capability to leverage informative modalities and cross-modal interactions across heterogeneous modality sets. \textsc{MuViS} is related to, but distinct from, several established problem classes. It is not generic TSER \cite{tan_monash_2020} or continuous sequence-to-sequence prediction \cite{bhirangi_hierarchical_2024}: \textsc{MuViS} assumes multimodal inputs and a physically grounded, sensor-based data-generating process, with targets representing hard-to-measure sensor variables rather than arbitrary output sequences. Moreover, unlike multimodal dynamic time-series learning \cite{mohapatra_maestro_2025}, \textsc{MuViS} focuses on continuous regression targets that mimic sensor measurements at a reference time, rather than discrete labels.

\section{Datasets}
\label{sec:datasets}

\textsc{MuViS} aggregates six benchmark datasets that span environmental monitoring, health sensing, vehicle dynamics, tire thermodynamics, chemical process monitoring, and electrochemical energy systems (Fig.~\ref{fig_eyecatcher}). Despite their heterogeneity, all tasks instantiate the \textsc{MuViS} formulation in Sec.~\ref{sec:muvis}: each training example is a multimodal sensor history window with a feature dimension $D$ and a sequence length $T$, and the target is a continuous, sensor-like scalar anchored at a reference time. In this work, we compare different baselines under the nominal case of fixed sequence length across modalities within each domain and their constant availability. We standardize each of the datasets, described in the following subsections, into fixed-length windows (done via resampling/aligning multi-rate channels) and evaluate models on consistent train/test splits. 

\subsection{Particulate Matter Estimation}
\label{sec:beijing}

The Beijing Multi-Site Air Quality data, introduced by Zhang et al. \cite{zhang_cautionary_2017}, is an environmental virtual sensing task, where the target is particulate matter concentration and inputs combine pollutant and meteorological measurements. The raw dataset includes measurements for air pollutants (SO$_2$, NO$_2$, CO, and O$_3$) and meteorological parameters, including temperature ($^\circ$C), pressure (hPa), dew-point temperature ($^\circ$C), rainfall (mm), and wind speed (m/s). Following TSER benchmark \cite{tan_monash_2020} constructions, we use the BeijingPM25Quality and BeijingPM10Quality reformulations, where each instance is a 9-dimensional daily multimodal time series of fixed length $T{=}24$ (hourly steps) and the continuous target is the corresponding PM$_{2.5}$ or PM$_{10}$ level. 

\subsection{Lateral Velocity Estimation}
\label{sec:revs}

\textsc{MuViS} includes the Revs Program Vehicle Dynamics Database, a public collection of instrumented vintage race-car data recorded during live events. The database integrates multimodal sensing sources (e.g., driver inputs, wheel/chassis measurements, and GNSS-aided inertial navigation), with signal-dependent sampling rates reported from 100\,Hz up to 1000\,Hz, and is distributed via the Stanford Digital Repository under an open data license \cite{kegelman_insights_2017}.
Following Brandt et al. \cite{brandt_faults_2025}, we define a virtual sensing task of estimating the vehicle's lateral velocity $v_y$, a key state for stability assessment and control that is typically not directly available from low-cost on-board sensing and therefore often inferred from other signals. In \textsc{MuViS}, we construct fixed-length windows with $D{=}12$ selected input channels, resample to a uniform rate (20\,Hz), and use short contexts ($T{=}20$) to reflect the fast-reacting nature of lateral dynamics near the friction limits.

\subsection{Tire Temperature Estimation}
\label{sec:tiretemp}

\textsc{MuViS} further includes a high-performance automated driving dataset that records vehicle state from RTK-GPS alongside control inputs, actuator states, and real-time tire temperature \cite{mori_vehicle_2025}. Despite being domain-wise similar to the Revs Database (Sec.~\ref{sec:revs}), this task targets the tire temperature $t_{\mathrm{tire}}$ collected using a Bridgestone POTENZA~S001. An Izze-Racing 16-channel infrared tire sensor measured tread temperature at the top of the tire and the target value is the mean of the middle 8 channels. $t_{\mathrm{tire}}$ is strongly coupled to traction and to rolling resistance effects relevant for vehicle efficiency and range \cite{tevell_creating_nodate}. In \textsc{MuViS}, we use $D{=}11$ vehicle-motion and control channels sampled at 1\,Hz and predict a single $t_{\mathrm{tire}}$ from a $T{=}50$\, temporal context, without using $t_{\mathrm{tire}}$ of the other three. We estimate the front left tire, as the front tires are of greater interest and present a greater challenge due to the steering and front-wheel drive of the car (the choice of the right tire is arbitrary). 

\subsection{Chemical Concentration estimation}
\label{sec:tep}

To cover industrial process monitoring, we use the Tennessee Eastman Process (TEP), a canonical simulated plant comprising a reactor, condenser, separator, recycle compressor, and product stripper \cite{downs_plant-wide_1993}.
The process defines a large set of measured variables and manipulated variables, and TEP data is widely used to evaluate inferential modeling when composition-related measurements are delayed or expensive. We use the widely adopted TEP simulation dataset \cite{rieth_additional_2017} sampled every 3 minutes, with 500 simulation runs (distinct random seeds); training trajectories contain 500 samples (25\,h) per run and testing trajectories 960 samples (48\,h) per run.
Following Ma et al. \cite{ma_soft_2024}, our \textsc{MuViS} task is the real-time estimation of one chemical's concentration, and we construct input windows from $D{=}33$ process channels (22 measurements and 11 manipulated variables). TEP is the only \textsc{MuViS} dataset for which no standard sequence length is established in the literature; based on preliminary experiments, we set the context to $T{=}20$ (i.e., 1\,h of history at 3-minute sampling). 

\subsection{Battery State-of-Charge}
\label{sec:soc}
Battery state-of-charge (SoC) estimation is a prototypical virtual sensing problem. It cannot be measured directly and must be inferred from measurable quantities such as current, voltage and temperature. We use the Panasonic 18650PF Li-ion dataset \cite{kollmeyer_panasonic_2018}, collected in a controlled lab environment and released as a public reference for comparing SoC estimation algorithms across standardized drive cycles and temperature conditions.
Prior work \cite{mondal_estimating_2024} reports this dataset at 10\,Hz sampling and includes experiments across multiple ambient temperatures (from $-20^\circ$C to $25^\circ$C), which induces substantial nonstationarity in voltage-current dynamics. Following de Mondal et al. \cite{mondal_estimating_2024}, in our MuVis benchmark, we use $D{=}7$ inputs (three primary measurements and four engineered channels), and we adopt a 12\,s context window ($T{=}120$ at 10\,Hz), matching common history sizes \cite{kollmeyer_panasonic_2018} used in recent learning-based SoC estimators.

\subsection{Heart Rate Estimation}
\label{sec:ppgdalia}
To represent health-related sensing under motion artifacts, \textsc{MuViS} includes the PPG-DaLiA \cite{attila_reiss_ppg-dalia_2019} dataset, recorded from 15 subjects using a chest-worn RespiBAN device (ECG ground truth) and a wrist-worn Empatica E4 device (BVP/PPG, EDA, temperature, and wrist acceleration). The raw data is multi-rate (e.g., BVP at 64\,Hz and wrist acceleration at 32\,Hz) and covers diverse daily activities, making it a challenging multimodal inference setting. Following established benchmark preprocessing  \cite{reiss_deep_2019}, we segment synchronized windows and define the target as heart rate (BPM) derived from ECG. In the course of our benchmark, we resample wrist channels to a uniform 64\,Hz grid, use $D{=}6$ wrist inputs (BVP, EDA, temperature, and 3-axis acceleration), and form fixed windows with $T{=}512$ (8\,s context).

\begin{table*}[t]
\caption{Predictive performance comparison across virtual sensing tasks. Results report Mean RMSE with 95\% bootstrap confidence intervals $[\text{Lower}, \text{Upper}]$ obtained from 200 iterations. Bold values indicate the best performing model per dataset.}
\begin{center}
\setlength{\tabcolsep}{3.5pt} % Adjusts padding to ensure width fits
\resizebox{\textwidth}{!}{
\begin{tabular}{l l l l l l l}
\toprule
\textbf{Dataset} & \textbf{XGBoost} & \textbf{CatBoost} & \textbf{MLP} & \textbf{ResNet1D} & \textbf{LSTM} & \textbf{Transformer} \\
\midrule
Lat. Velocity: Monterey & $0.120_{[0.118, 0.122]}$ & $0.109_{[0.107, 0.110]}$ & $0.131_{[0.129, 0.133]}$ & $0.110_{[0.109, 0.112]}$ & $\mathbf{0.108_{[0.107, 0.110]}}$ & $0.119_{[0.117, 0.120]}$ \\
\midrule
Lat. Velocity: Targa '13 & $0.098_{[0.095, 0.100]}$ & $0.087_{[0.085, 0.090]}$ & $0.082_{[0.080, 0.085]}$ & $0.071_{[0.069, 0.073]}$ & $\mathbf{0.070_{[0.068, 0.072]}}$ & $0.103_{[0.100, 0.107]}$ \\
\midrule
Lat. Velocity: Targa '14 & $\mathbf{0.077_{[0.076, 0.078]}}$ & $0.080_{[0.079, 0.081]}$ & $0.089_{[0.088, 0.091]}$ & $0.079_{[0.078, 0.081]}$ & $0.078_{[0.077, 0.080]}$ & $0.099_{[0.097, 0.100]}$ \\
\midrule
PM 10 & $\mathbf{91.642_{[86.170, 98.755]}}$ & $92.091_{[86.628, 99.403]}$ & $95.889_{[90.846, 102.116]}$ & $95.928_{[90.383, 102.590]}$ & $99.291_{[93.251, 106.685]}$ & $103.852_{[97.889, 110.958]}$ \\
\midrule
PM 2.5 & $60.983_{[55.684, 66.646]}$ & $61.366_{[56.293, 66.877]}$ & $60.477_{[55.777, 65.810]}$ & $62.994_{[58.662, 67.944]}$ & $\mathbf{59.726_{[55.845, 64.412]}}$ & $65.600_{[60.266, 71.800]}$ \\
\midrule
Tire Temp. & $3.758_{[3.516, 3.968]}$ & $3.840_{[3.606, 4.071]}$ & $3.173_{[2.943, 3.437]}$ & $3.401_{[3.184, 3.590]}$ & $\mathbf{2.447_{[2.298, 2.608]}}$ & $3.653_{[3.372, 3.987]}$ \\
\midrule
Chemical Conc. & $\mathbf{0.051_{[0.051, 0.051]}}$ & $\mathbf{0.051_{[0.051, 0.051]}}$ & $\mathbf{0.051_{[0.051, 0.052]}}$ & $0.058_{[0.058, 0.058]}$ & $0.061_{[0.061, 0.062]}$ & $0.058_{[0.057, 0.058]}$ \\
\midrule
Battery SoC & $0.025_{[0.025, 0.025]}$ & $0.011_{[0.011, 0.011]}$ & $0.008_{[0.008, 0.008]}$ & $\mathbf{0.007_{[0.007, 0.007]}}$ & $0.008_{[0.008, 0.008]}$ & $0.009_{[0.009, 0.009]}$ \\
\midrule
Heart Rate & $9.526_{[9.327, 9.747]}$ & $9.035_{[8.844, 9.214]}$ & $11.519_{[11.262, 11.806]}$ & $\mathbf{3.943_{[3.886, 4.013]}}$ & $8.528_{[8.355, 8.693]}$ & $5.242_{[5.070, 5.434]}$ \\
\bottomrule
\end{tabular}
}
\label{tab:rmse_results_detailed}
\end{center}
\vspace{-17pt}
\end{table*}

\section{Baselines and Evaluation Protocol}
\label{sec:Baselines}

\begin{figure}[t]
\centerline{\includegraphics[width=0.5\textwidth]{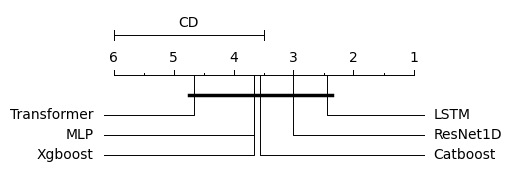}}
\caption{Critical distance diagram. The non-significant Friedman test indicates no statistically superior architecture.}
\label{fig:cd}
\vspace{-17pt}
\end{figure}

We evaluate six baselines spanning diverse inductive biases. Gradient-boosted decision trees (GBDTs), specifically XGBoost \cite{chen_xgboost_2016} and CatBoost \cite{prokhorenkova_catboost_2019}, are compared against four deep learning architectures: a Multi-Layer Perceptron (MLP), an LSTM \cite{hochreiter_long_1997}, a ResNet 1D-CNN and a BERT-like Transformer encoder with learnable positional encodings \cite{devlin_bert_2019}. For non-temporal models (GBDTs and MLP), we utilize sequence-flattening to incorporate temporal context into the feature space. Rigorous hyperparameter optimization is performed using Optuna with a Tree-structured Parzen sampler over 100 trials per model-dataset pair \cite{bergstra_algorithms_2011}. We optimize for Root Mean Squared Error (RMSE) on a 10\% validation split and report final performance on a hold-out test set. We report results via 200-fold bootstrapping of the test set. All experiments were executed on a single H100. As shown in Tab.~\ref{tab:rmse_results_detailed}, performance is highly domain-dependent. GBDTs demonstrate remarkable robustness, achieving state-of-the-art results on the \textit{Vehicle Dynamics}, \textit{Tennessee Eastman}, and \textit{Beijing PM10} datasets. Among NN architectures, LSTMs and ResNet1D prove most effective for the \textit{REVS} racing datasets and \textit{Beijing PM2.5}. A statistical comparison using the Friedman test did not find evidence of a statistically significant difference ($p = 0.209$). As shown in Fig.~\ref{fig:cd}, the rank difference between the highest and lowest-ranked models remains within the Nemenyi critical distance ($CD = 2.513$), indicating no significant differences between models \cite{herbold_autorank_2020}. These results suggest that standard deep learning and tree-based methods reach a performance plateau in virtual sensing. The lack of a significant winner underscores the need for specialized architectures designed to move beyond the limitations of generic models and provide consistent, generalized improvements in virtual sensing tasks.

\section{Conclusion and Future Work}

\label{sec:Conclusion}
We introduced \textsc{MuViS}, a domain-agnostic benchmark designed to standardize multimodal virtual sensing evaluation through diverse datasets and representative architectures. Our evaluation compares carefully tuned baselines, demonstrating that while gradient-boosted ensembles remain highly competitive, the landscape is nuanced, with specific NN architectures excelling in distinct domains. By providing these standard benchmarks, \textsc{MuViS} enables the rigorous assessment of novel architectures within a unified framework. Future research should address the inherent complexities of real-world deployment, specifically dynamic sensor availability and handling malfunctions, heterogeneous sampling rates, and imbalanced regression in target distributions. We intend to continuously expand this open-source archive and invite community collaboration to foster the development of universal, robust virtual sensing models for complex engineering systems. Ultimately, \textsc{MuViS} advances the connection between real-world processes and their digital counterparts by enabling more reliable, data-driven state estimation.

\section*{Acknowledgment}
This work is funded by the European Commission Key Digital Technologies Joint Undertaking - Research and Innovation (HORIZONKDT-JU-2023-2-RIA), under grant agreement No 101139996, the ShapeFuture project - "Shaping the Future of EU Electronic Components and Systems for Automotive Applications".

\bibliographystyle{IEEEtran}
\bibliography{VSB_cleaned}

\end{document}